# Size Effect in Electron Paramagnetic Resonance Spectra of Impurity Centers in Diamond Nanoparticles

G. G. Zegrya[1,*], D. M. Samosvat[1], V. Yu. Osipov[1], A. Ya. Vul'[1], and A. I. Shames[2]

[1] Ioffe Institute, 26 Politekhnicheskaya st., St. Petersburg, 194021 Russia

[2] Department of Physics, Ben-Gurion University of the Negev, Beer-Sheva, 8410501 Israel

The evolution of the polycrystalline pattern of electron paramagnetic resonance (EPR) spectra of intrinsic and induced paramagnetic centers in an ensemble of submicrometer diamond particles on diminishing average particle size is considered. Recent experimental data unambiguously demonstrate consistent reduction and then zeroing of typical hyperfine pattern from P1 centers on approaching a certain particle size and below it. These changes are accompanied by appearance and strengthening of new structureless signals. In small particles the electron wave function of a surface paramagnetic center is delocalized over the whole nanoparticle. In result the electron spin "experiences" the average field of all surrounding nuclei, which is zero. At the same time, the paramagnetic centers localized in the bulk of a nanoparticle are also cut off the hyperfine interaction due to the electron spin diffusion. Thus, upon the nanodiamond particle size decreases the hyperfine related features of the polycrystalline EPR spectrum become weaker and the spectrum appear to be structureless.

# 1. Introduction

Last decades demonstrated explosive growth of technologies concerned with a variety of nanomaterials. Among all these developments nanosized carbon particles and, especially, nanodiamonds are in the first flight.[1] The reason for such an interest just to diamond nanoparticles is associated with both already implemented and progressively developing applications of nanodiamonds in various fields: from biomedicine as fluorescent labels and carriers for targeted drug delivery, to spintronics and quantum computing.[2-4] In the view of these applications the very important objective is fabrication of diamond nanoparticles having sizes within the range of 2–10 nm and controlled content of negatively charged nitrogen-vacancy centers (NV⁻). The latter " triplet color centers" being both paramagnetic and optically active are responsible for the application of nanodiamonds as fluorescence labels in biomedical applications,[5] extremely high sensitive magnetic field/gradients sensors,[6] and qubits implemented at room temperatures.[7]

In the simplest case, a nitrogen atom plays role of a substitutional impurity in the diamond lattice, with each of its four valence electrons forming covalent bonds between carbon atom and nitrogen, and the fifth, with an uncoupled electron orbital determining the paramagnetic properties of the center. Such a center with electron spin $S = 1/2$, known as P1 center, is in the neutral charge state. This center is well identified by electron paramagnetic resonance (EPR). A characteristic feature of EPR spectra of P1 centers in both mono- and polycrystalline diamond samples is well resolved hyperfine structure (HFS) occurring due to the interaction of the uncoupled electron with the neighboring $^{14}$N nuclei ($I = 1$) (**Figure 1**): nearly equidistant signals are observed on both sides of the central signal characterized by $g = 2.0024$.

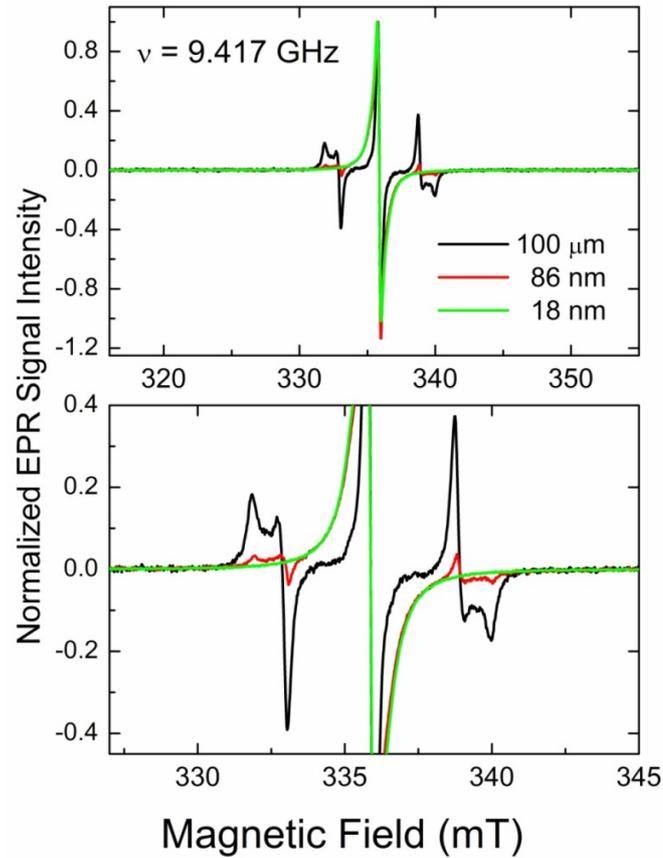

**Figure 1.** Room temperature X-band ($\nu$ = 9.417 GHz) EPR spectra of polycrystalline diamond samples with average particle sizes of 100 $\mu$m (black line), 86 nm (red line), and 18 nm (green line).

The aforementioned technological applications require fabrication of diamond nanocrystals having sufficiently high content of NV$^-$ centers.[9] It has been found that, as diamond nanocrystals become smaller, not only the content of NV$^-$ centers decreases, but also the shape of the polycrystalline EPR pattern of P1 centers undergoes drastic changes.[10] Thus, detailed studies[12–14] of the EPR spectra of diamond powders produced by mechanical grinding and subsequent fractionation (from 100 $\mu$m to several tens of nanometers) revealed three distinctive modifications appearing in the EPR spectrum upon a decrease in the average size of particles in a polycrystalline sample:

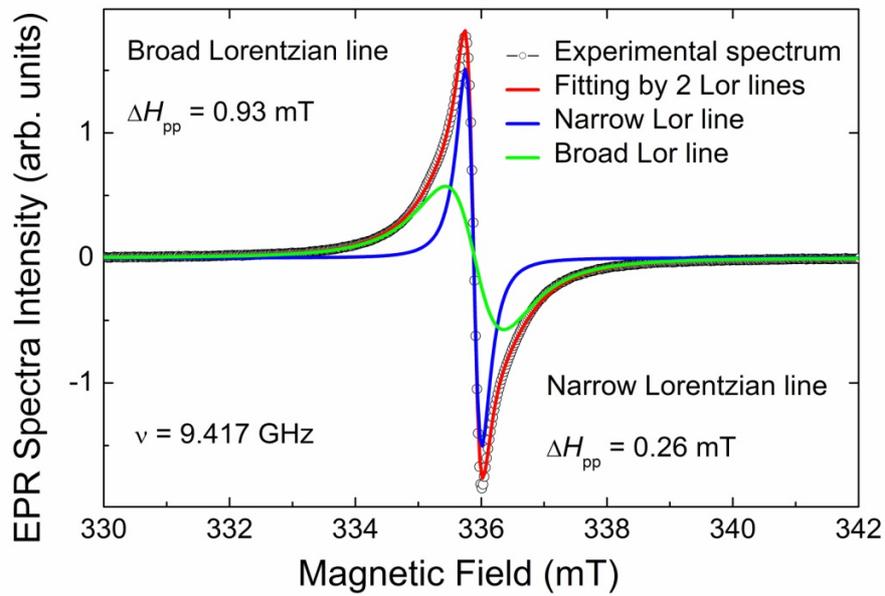

**Figure 2.** Room temperature X-band ($\nu$ = 9.417 GHz) EPR spectrum of fine nanodiamonds (average particle size 18 nm). The experimental spectrum (black open circles) was simulated as a supeposition of two Lorentzian lines (red trace) having the same *g*-factor (*g* = 2.0028) but different line widths: narrow (blue trace) and broad (green trace).

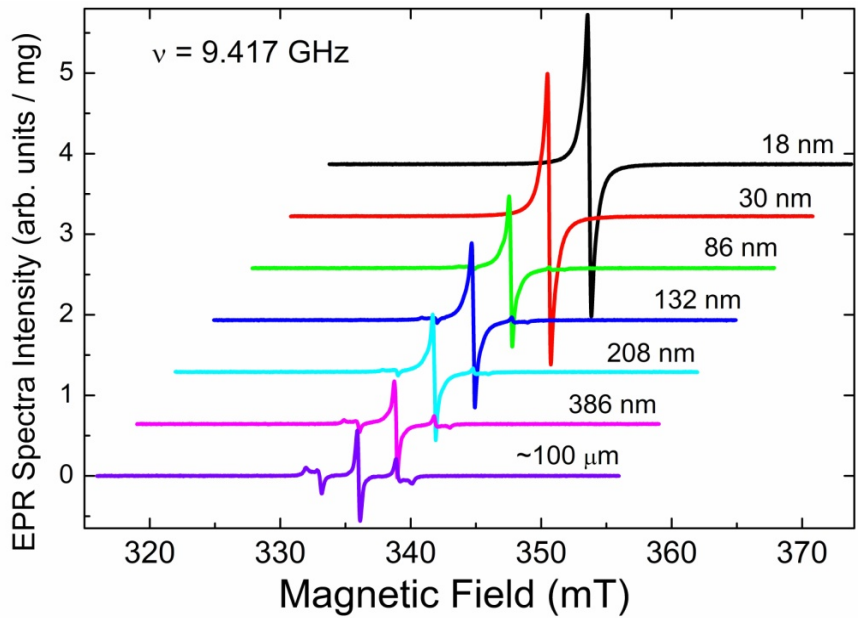

**Figure 3.** Consecutive changes of the EPR spectrum of polycrystalline diamonds with decreasing average particle size. The spectra were recorded in the X-band ($\nu$ = 9.417 GHz) at room temperature.[11]

- Steadily decreasing contribution of the typical P1-pattern with $g$ = 2.0024 and well resolved hyperfine structure to the EPR spectra observed.

- Increasing contribution of the broad singlet signal with g = 2.0028, associated with surface and near-surface defects induced by grinding (dangling bonds).

- Appearance of a new structureless narrow signal with $g$ = 2.0028 and further steadily increasing contribution of this signal to the resulting spectrum.[12,13] On the other hand, increase in size of diamond particles from nano- to micro-meters leads to restoration of the hyperfine split EPR pattern typical for P1 centers, which has not been observed in the starting 5 nm detonation nanodiamonds.[15]

In this study, we suggested a model that accounts a mechanism of modification of the P1 centers related EPR spectra of small diamond nanoparticles.

The main concepts of the model consist as follows.

On formation of a nanodiamond particle (both by mechanical grinding and by dynamic synthesis), some new paramagnetic centers, additional to the intrinsic ones, appear on the surface of a particle and within its subsurface interface layer. These new paramagnetic centers were attributed to dangling bonds[12,13] or, erroneously, to negatively charged vacancies.[14] Due to relatively high local densities and short distances, characteristic for a nanosized system, these new defects interact both with each other and neighboring P1 centers. It is supposed that just these interactions cause experimentally observed changes in the EPR spectrum: there appear two high-intensity singlet lines with Lorentzian shape with different half-widths ("broad" $\Delta H_{pp} = 0.93\ mT$ and "narrow" $\Delta H_{pp} = 0.26\ mT$) and practically coinciding resonance fields (*g*-factors).

The model suggests that the "broad" line, being absent (or negligible) in EPR spectra of coarse micrometer-sized diamond particles, has structureless shape due to interactions between defects in the narrow subsurface layer.

As shown below, the shape of the "broad" line is a result of the delocalization of the electron wave function of surface and subsurface centers throughout the nanoparticle. At the same time, two "narrow" lines, observed in the EPR spectra of smaller diamond nanoparticles, are of unlike nature: the line with $\Delta H_{pp} \approx 0.16\ mT$ appears originates from non- (or weakly-) interacting P1 centers located in cores of coarser diamond micro- and nanoparticles, whereas the line with $\Delta H_{pp} \approx 0.26\ mT$ appears as a result of the exchange interaction between these P1 centers and delocalized surface/subsurface spins occurring in smaller particles.

All the three above-described lines are well observed in the spectra of nanocrystals with intermediate average particle sizes. All the HFS-split EPR signals from localized P1 centers fully disappear due to the spin diffusion induced by the indirect exchange interaction between bulk centers via delocalized electrons of surface centers, because an electron in a bulk paramagnetic center "experiences" the averaged field of all the nuclei, which is zero (see below).

## 2. Model and general concepts

The model suggested is based on the assumption that a change in the EPR spectrum of paramagnetic centers occurs as a particle becomes smaller because of the strong interactions between the centers in the bulk and in the thin subsurface layer $4\pi R^2 \Delta R$, where $\Delta R$ is the distance from the nanodiamond surface within which a center is considered a surface center, and $R$ is the particle radius (see **Figure 4**). Such a surface center will be accurately defined below.

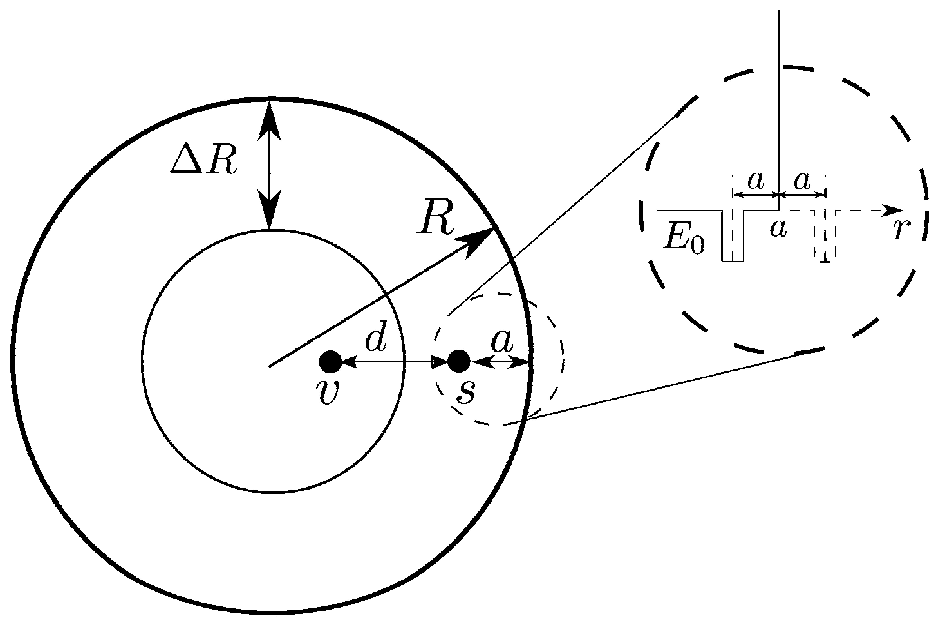

**Figure 4.** Model and formulation of the problem. The figure shows the relative positions of the paramagnetic centers under consideration. A diamond nanoparticle of radius R and a surface center (*s*) located in the subsurface layer at a distance (*a*) from the surface are considered. The bulk center (*v*) is located at a distance *d* from the surface center *s*. The inset shows a paramagnetic center at a distance *a* from the diamond surface approximated with an infinitely high barrier. The dashed circle shows the image of the center.

It is known[16] that the main mechanisms responsible for the interaction between the electron spins of the paramagnetic centers are the spin-spin (dipole) and exchange mechanisms. As the particle size decreases, the role of the exchange interaction between the bulk and surface centers becomes more important. In addition, an interaction between the spins of electrons with the magnetic moments of their own or neighboring nuclei exists in the system (hyperfine interaction). This interaction is responsible for the appearance of the hyperfine structure (HFS) in the EPR spectra. The energies of the spin-spin and exchange interactions differently depend on distances between the interacting electron spins and on the electronic structure of a substance. Starting at a certain radius of a nanodiamond particle (i.e., at a certain critical nanoparticle radius $R_{cr}$), the direct exchange interaction becomes stronger than the hyperfine one. Moreover, and this is high importance, surface/subsurface paramagnetic centers induced by grinding start to introduce increasing contribution to the formation of the observed EPR spectrum as the nanoparticle size decreases. For instance, surface paramagnetic centers enhance the spin diffusion between bulk centers due to the appearance of an indirect exchange interaction between the bulk centers (**Figure 5**). This leads to averaging of the hyperfine interaction and, correspondingly, to disappearance of the satellite HFS lines in the spectrum of bulk paramagnetic centers.

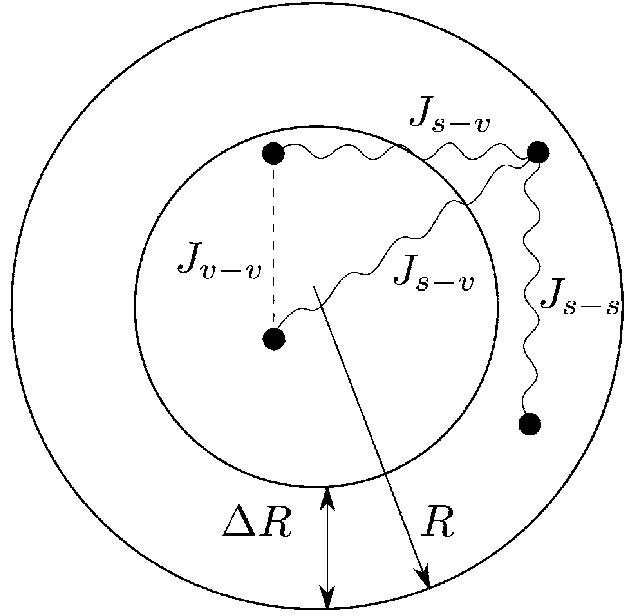

**Figure 5.** Indirect exchange interaction between the bulk centers via delocalized states of the surface centers. $J_{v-v}$ is the energy of the indirect exchange interaction between the bulk centers, $J_{s-v}$ is the exchange interaction between the bulk and surface centers, and $J_{s-s}$ is the energy of interaction between the surface centers.

By the model suggested, when considering mechanism causing the radical change of the EPR spectrum upon a decrease in the particle size, it is necessary to compare the matrix element of the dipole-dipole (spin-spin) interaction between the bulk and surface paramagnetic centers, $M_{DD}(R_{cr})$, and the matrix element of the exchange interaction between the bulk and surface centers, $J_{s-v}(R)$, as well as indirect exchange interaction between the bulk centers, $J_{v-v}(R)$, with characteristic hyperfine splitting energies.

The equality

$$\Delta E_{hyperfine} = J_{v-v}(R_{cr}) \qquad (1)$$

will be used to find the critical diameter $2R_{cr}$ of nanodiamond from which the indirect exchange interaction starts to suppress manifestations of the hyperfine splitting. Further it will be shown that the energy of the dipole-dipole interaction is two orders of magnitude lower than the energy of the hyperfine splitting energy, and, therefore, it cannot be responsible for the disappearance of the characteristic HFS in the EPR spectrum of P1 centers.

It is known that the binding energy of an electron at an impurity center may change significantly when the impurity is close to the crystal surface.[17–19] It is noteworthy that the indirect exchange interaction plays a key role in explaining the effect of spin diffusion between the bulk centers. Indeed, for the P1 centers lying outside the subsurface layer, the direct exchange interaction between the bulk centers does not suppress the HFS. Nevertheless, these centers, as well, are not involved in the HFS because of the spin-diffusion mechanism due to the indirect exchange interaction via the surface centers.

Let us consider in detail this indirect exchange interaction mechanism.

Let a paramagnetic center be located at a distance $a$ from the diamond surface, which, for simplicity, is represented as an infinite potential barrier (**Figure 4**). The dashed line in the inset is the mirror image of the center.[17]

The typical distance from the nanodiamond boundary, $\Delta R$, at which the surface effect becomes significant, is determined by the characteristic scale of the decay of the localized state wave function, $\kappa_v^{-1}$:

$$\Delta R \simeq \kappa_v^{-1} = \frac{\hbar}{\sqrt{2mE_0}} \qquad (2)$$

where $m$ is the effective electron mass at a center, and $E_0$ is the electron binding energy at a center. We consider paramagnetic centers localized in a spherical layer with thickness $\Delta R$ to be of the surface type.

It is important that the spectrum and wave functions of surface paramagnetic centers strongly differ from the spectrum and wave function of bulk centers (**Figure 4**). The binding energy of an electron on a surface center is lower than that for the bulk center: the "tail" of the wave function for the surface center, $\kappa_s^{-1}$, is substantially longer than that for the bulk center $\kappa_v^{-1}$.

To find the coordinate part of the electron wave function at a paramagnetic center, we use the model of the zero-radius potential.[16] It will be recalled that the model of the zero-radius potential is applicable to describe the wave function of an electron localized at a deep center and is based on representing the potential of the center as a delta-function quantum well.[16]

The possibility of applying the model of the zero-radius potential to deep centers is based on that the radius at which the impurity potential is effective, $\lambda_0$, is on the order of the lattice constant, and the extent of the wave function $\kappa_v^{-1} = \hbar/\sqrt{2mE_0}$ for the deep center is several times the lattice constant. This is justified when the binding energy of an electron at a paramagnetic center, $E_0$, is lower than the energy gap width of nanodiamond, $E_g$, and the electronmass $m$ is smaller than the free electron mass $m_0$.

For diamond, the corresponding parameters have the following values: $E_g = 5.45$ eV and $E_0 = 1.7$ eV.

As shown below, the main effect leading to a change of the EPR spectrum of a nanodiamond particle is the indirect exchange interaction, both between the bulk centers and between the surface and bulk centers. This interaction is introduced by analogy with the indirect exchange interaction (RKKY) (Ruderman–Kittel– Kasuya–Yosida) between magnetic atoms in ferromagnetics[20] and in quantum dots,[21], with role of a "transfer agent" played in this case by delocalized electrons of surface centers, instead of the conductivity electrons.[22]

The indirect exchange interaction between bulk centers becomes possible due to the delocalization of the wave functions of the surface centers throughout the nanodiamond volume. It is noteworthy here that a precise knowledge of the nature of surface paramagnetic centers is by no means necessary. As surface centers that can be involved in the effect of indirect exchange interaction can also serve P1 centers in the near-surface layer and dangling bonds on the nanodiamond surface induced by the grinding. It is noteworthy that the dipole-dipole interaction in the system under consideration cannot take part in the disappearance of the HFS spectrum because, by its nature, the energy of such an interaction depends on the distance d between the centers as $1/d^6$. At typical distances between the paramagnetic centers in a diamond nanoparticle, the energy of the dipole-dipole interaction is low as compared with the hyperfine interaction energy.

As shown in,[22] the wave function of the surface paramagnetic centers is partly delocalized in the nanodiamond volume, in agreement with qualitative concepts. In addition, bulk centers are bound to each other, due to the effect of indirect exchange interaction, which results, because of the spin diffusion effect, in the averaging of the effective hyperfine splitting at nitrogen nuclei. Just this circumstance leads to the disappearance of the HFS in the EPR spectra of P1 centers in fine particles.

# 3. Energy spectrum and wave functions of paramagnetic centers

## 3.1. Bulk Paramagnetic Centers

To calculate the probability of interaction between the bulk and surface centers, it is necessary to know the wave functions of electrons in the spin states $\pm 1/2$. In the simplest case, an electron at an impurity center is described by a wave function in the form of a product of the coordinate part by that of the spin type:

$$\Psi(\xi) = \phi_k(r)\chi_\sigma(s) \qquad (3)$$

where $\xi = (r, \sigma)$, $\sigma$ is the spin quantum number, $r$ is the coordinate part of the wave function and $\chi_\sigma(s)$ is the spin part. We further consider the wave functions of the paramagnetic centers in terms of the zero-radius potential model.[16,17]

In this case, the electron wave function is mostly concentrated in the range where the potential of the atom is zero.

For the ground (s) state, we have

$$\phi(r) = C \frac{exp\left(-\kappa_v r\right)}{r}, \qquad (4)$$

$$\kappa_v = \frac{\sqrt{2mE_0}}{\hbar}, \qquad r > \lambda_0, \qquad (5)$$

where $\lambda_0$ is the range of action of the potential, and $E_0$ is the binding energy of an electron at the impurity. The constant $C$ is found from the normalization condition:

$$\int |\phi(\mathbf{r})|^2 \, d^3r = 1. \tag{6}$$

If $\kappa_v \lambda_0 \ll 1$ then $C = \sqrt{\frac{\kappa_v}{2\pi}}$, if $\kappa_v \lambda_0$ is not too small then, $C = \sqrt{\frac{\kappa_v}{2\pi}} \frac{1}{1+\beta}$, where $\beta$ is the parameter characterizing the contribution of the region $r < \lambda_0$ to the normalization integral.[17]

## 3.2. Surface Paramagnetic Centers

Let us consider how the electron wave function of a paramagnetic center behaves near the surface of a diamond particle and how the electron energy at the impurity center changes.

With consideration for the image potential (see **Figure 4**), the wave function of a surface center has the form

$$\phi(r) = A \left( \frac{exp\,(-\kappa_s r)}{r} - \frac{exp\,(-\kappa_s |\mathbf{r}-2\mathbf{a}|)}{|\mathbf{r}-2\mathbf{a}|} \right) \tag{7}$$

The quantity $\kappa_s$ is related to the binding energy $E$ of an electron at a center situated close to the surface by

$$E = \frac{\hbar^2 \kappa_s^2}{2m} \tag{8}$$

Substituting $\phi_s(r)$ from (7) into the boundary conditions for the wave function,[17] we obtain the equation for finding $\kappa$ and, accordingly, the energy $E$:[17]

$$\kappa_s + \frac{exp\,(-2\kappa_s a)}{2a} = \kappa_v. \tag{9}$$

For an impurity center lying far from the surface, when .$va \gg 1$, the boundary hardly affects the electron level position of the paramagnetic center, with the electron binding energy decreasing.

Equation (9) describes how the spectrum of an impurity center varies when it approaches the surface (boundary) of a nanoparticle, with the electron binding energy decreasing. At $a = (1/2)\kappa_v$, $\kappa_s = 0$ and the electron level goes to the continuous spectrum.

Thus, we name surface paramagnetic centers those centers that fall within a spherical layer with thickness $\Delta R = 1/\kappa_v$, where $\kappa_v$ is found from the energy at which the bulk center lies.

## 4. Matrix Elements of the Main Interactions in Nanodiamond

As already noted, the specific structure and shape of the EPR spectrum in a diamond nanoparticle are determined by three main interactions: dipole-dipole (or spin-spin), exchange interaction between electrons of paramagnetic centers, and hyperfine splitting associated with the interaction between spins of nuclei and spins of electrons.[16] The latter leads to a well-resolved HFS of the EPR spectra of coarse diamond particles, which is primarily characterized in the case of P1 centers by the presence of additional side satellite signals.

Let us discuss in detail each of these interactions and calculate their energy.

### 4.1. Matrix Element of the Dipole-Dipole (Spin-Spin) Interaction

The dipole-dipole interaction is responsible for the interaction between the magnetic moments of unpaired electrons of the centers and, consequently, it is proportional to $\sim \frac{\mu_1 \mu_2}{d^3}$ where $\mu_1$ and $\mu_2$ are the magnetic dipole moments.

The precise expression for the matrix element of the spin-spin interaction is given in Appendix A.

In the end, the matrix element of the spin-spin interaction is given by

$$M_{s-s} = \frac{1}{2}\left(\frac{e\hbar}{mc}\right)^2 \frac{1}{d^3} \tag{10}$$

The contribution from this matrix element is numerically estimated in Section 4.4.

## 4.2. Matrix Element of the Exchange Interaction

Let us now consider the matrix element of the exchange interaction. The exchange interaction is of nonmagnetic nature and is associated with the Coulomb interaction between electrons from different centers. Accordingly, the matrix element of the Coulomb interaction is given by

$$M_c = \langle \psi_1 | \frac{e^2}{\varepsilon |\boldsymbol{d} + r_1 - r_2|} | \psi_2 \rangle \tag{11}$$

Here, $K$ is the Coulomb integral, and $J$ is the exchange integral.

By the virtue of the spin selection rules, the matrix element of the direct Coulomb interaction $K$ is zero, and the nonzero matrix element corresponds to the exchange interaction $J$. For convenience of calculation, we pass to the coordinates reckoned from the first center.

It can be shown that that the matrix element of the exchange interaction between bulk centers has the form:

$$J_{s-v} = \frac{5}{4} \frac{e^2}{\varepsilon} \frac{1}{\kappa^3 V} \frac{\kappa_v}{2\pi} \tag{12}$$

where $V$ is the volume of a diamond nanoparticle.

## *4.3. Indirect Exchange Interaction*

Also possible in our system is the indirect exchange interaction between bulk paramagnetic centers due to their interaction via unpaired electrons of delocalized surface centers (**Figure 5**). According to the Classical RKKY theory,[20] this process is described in terms of the second order of the perturbation theory.

According to the perturbation theory, the energy of this interaction is on the order $J_{v-v} \sim \frac{J_{s-v}^2}{\Delta E} \approx \frac{e^2}{\varepsilon} \frac{1}{\kappa_v^4 R^5}$, where $Js - v$ is the energy of interaction between a surface and a bulkcenters. Here, the characteristic energy $\Delta E$ is the line halfwidth of a surface paramagnetic center, associated with the overlapping of the wave functions of the surface centers. The interaction between the surface centers can give rise to a mini-band with a characteristic width on the order of this interaction in the energy spectrum. This enables an indirect exchange interaction by analogy with conductivity electrons in a metal, which are intermediaries in this interaction.

## *4.4. Hyperfine Splitting*

In addition to the spin-spin and exchange interactions, the system under consideration shows a hyperfine splitting describing the interaction between the magnetic moment of a nucleus and an unpaired electron at the paramagnetic center. This interaction leads to an additional splitting of EPR lines (HFS).

We believe that the electron wave function at a surface center is delocalized throughout the volume of a diamond nanoparticle (see (26)).[18] The nuclear spins of paramagnetic centers in nitrogen are situated chaotically because the energy $k_B T$ substantially exceeds the hyperfine splitting and the nuclear Zeeman interaction. It will be recalled that the Hamiltonian of the interaction of a spin with the magnetic field in the presence of a hyperfine splitting has the form

$$H = A(\mathbf{sI}) + g_e\mu_B(\mathbf{sH}) \qquad (13)$$

Here, $A$ is the constant of hyperfine splitting dependent on the configuration of the wave functions of the electron and nucleus, $I$ is the magnetic moment of the nucleus, $g_e$ is the $g$-factor of an electron at the center, and $\mu_B$ is Bohr's magneton. In the case of a delocalized function of the paramagnetic center, the electron experiences not only the magnetic field of that nucleus in the field of attraction of which it is situated, but also a certain effective field created by all nitrogen nuclei in the system under consideration. This is so because the spin-spin interaction strongly decreases with increasing distance, and the electron wave function at a surface enter is delocalized, which means a nearly "contact" interaction of the delocalized wave function with nitrogen nuclei in the system under consideration. In this case, the effective magnetic field experienced by the electron is given by

$$\delta H_N = \sum_n A_n \mathbf{I} \frac{1}{g_e\mu_B} \qquad (14)$$

On being averaged over all nuclei, the magnetic field $\langle\delta H_N\rangle = 0$ because the average value of the spin of nuclei is zero due to their being chaotic.[23]

## 4.5. Spin Diffusion

The spin diffusion is the process of a spatial leveling of the nonuniform spin polarization in the system of localized magnetic moments. In contrast to the ordinary diffusion, the spin diffusion results only in the leveling-off of the polarization, without a mass transfer. In this case, the leveling of the spin polarization is described by the ordinary diffusion equation. The spin diffusion process occurs due to the spin–spin interaction and exchange interaction. In our system, the dipole-dipole (or spin-spin) interaction can be disregarded because it is two orders of magnitude weaker than the exchange interaction between bulk and surface centers. The spin diffusion frequency is determined in terms of the interaction energy as $\nu = \frac{J}{2\pi\hbar}$, where $J$ is the matrix element of the interaction between the centers. In our case, the spin diffusion coefficient is found as $D = \frac{\frac{1}{3}\langle d_{ij}^2\rangle\langle J\rangle}{\hbar}$, where $\langle d_{ij}^2\rangle$ is the averaged squared distance between the centers. If the spin diffusion frequency $\nu$ exceeds the hyperfine splitting frequency, the HFS lines converge toward the center.

## *4.6. Numerical Estimates for the Matrix Elements of Interactions in Nanodiamonds of Various Sizes and Comparison with Experimental Data*

Let us now present general expressions for the characteristic energies of the spin–spin and exchange interactions. The energy associated with the spin–spin interaction is given by

$$\hbar\omega_{dd} = \frac{e^2\hbar}{2m^2c^2}\frac{1}{d^3} \tag{15}$$

Here, $d$ is the distance between the paramagnetic centers.

The following expression can be written for the interaction of surface centers with bulk centers:

$$\hbar\omega_{s-v} = \frac{5}{4}\frac{e^2}{\varepsilon}\frac{1}{æ_v^2 V}\frac{1}{2\pi\hbar} \tag{16}$$

The matrix element of the direct exchange interaction between bulk centers is negligible.

We now present numerical estimates for various matrix elements. We take the following parameters: $d = 40$ Å (average distance between centers, which corresponds to their concentration of 150 ppm in the bulk material, according to experimental data), $m = 0.57 \cdot m_0$, $R_1 = 15$ nm (radius of fine particles with $R < R_{cr}$), $R_2 = 100$ μm (radius of coarse particles with $R > R_{cr}$), $E_b$=1.7 эB eV (binding energy of a deep center in the bulk material). Here and hereinafter, we give estimates for the matrix elements of the spin-spin and exchange interactions and the hyperfine splitting in the frequency range

$$\hbar\omega_{dd}^{(1)} = \hbar\omega_{dd}^{(2)} \sim 5 \cdot 10^{-9} eV$$

Here, the superscript (1) corresponds to fine particles, and (2), to coarse particles considered in the experiment.

For the exchange contribution, the matrix element between the bulk centers has an order of magnitude

$$\hbar\omega_{s-v}^{(1)} = \hbar\omega_{s-v}^{(2)} \sim 2{,}5 \cdot 10^{-42} eV$$

For the exchange contribution, the matrix element between the bulk and surface centers has the following order of magnitude

$$\hbar\omega_{s-v}^{(1)} \sim 2 \cdot 10^{-9} eV$$

$$\hbar\omega_{s-v}^{(2)} \sim 5 \cdot 10^{-18} eV$$

The cyclic frequency of the paramagnetic transition between states in the magnetic field, corresponding to the working frequency of 9.4 GHz:

$$\omega_{1-2} = 2\pi\nu = 5.91688 \cdot 10^{10} s^{-1}$$

The hyperfine splitting frequency (see **Figure 1**)

$$\hbar\omega_{\text{hf}} \sim 5.4 \cdot 10^{-7} eV$$

According to our estimate, the frequency of the indirect exchange interaction can be written as

$$\hbar\omega_{v-v} = \frac{e^2}{\varepsilon \varkappa^4 R^5} N_s,$$

where $N_s$ is the number of surface centers judged from their concentration of 0.2–0.3 spin per square nanometer.

For diamond particles with size smaller than the critical value, $R_1 = 15$ nm, we have $\hbar\omega_{v-v} \sim 10^{-7}$ eV.

For intermediate particle sizes $R_1<R<R_2=50$ nm), $\hbar\omega_{v-v} \sim 6 \cdot 10^{-8}$ eV. This estimate gives reason to state that $R_{cr} \sim 50$ nm.

For coarse particles with R> 50 nm, we have the following estimate:
$$\hbar\omega_{v-v}^{(2)} \sim 10^{-17} \, eV$$

The spin diffusion frequency at a strong exchange interaction is $\omega_{v-v}^{(1)}$. If the radius of a diamond nanoparticle is small, the exchange interaction between bulk and surface centers becomes equal to the hyperfine splitting, which, accordingly, leads to the averaging of the latter.

Thus, it can be seen that, in the conditions in which there are paramagnetic centers on the surface of a nanodiamond, with their wave function delocalized over the volume of the nanocrystal, the indirect exchange interaction between bulk centers becomes effective. This also gives rise to the spin diffusion between bulk centers.

The half-width of levels will be precisely calculated elsewhere on the basis of the density-matrix formalism.[24]

The above numerical estimates make it possible to fully describe the observed change in the EPR spectra of polycrystalline samples as result of the systematic decrease in the average size of diamond particles.

## 5. Conclusions

Thus, the analysis done in the study allows to suggesting a mechanism that causes changes in the shape of the EPR spectra of nitrogen impurity centers, observed upon the systematic decrease in the average size of diamond nanoparticles.

The main concepts on which this mechanism is based are the following: (i) the structure of the wave functions of a surface center strongly differs from that of the wave functions of a bulk center; (ii) surface centers are not involved in the hyperfine splitting because their wave function is delocalized throughout the nanoparticle and they "experience" the averaged field of all the nuclei, which is zero; (iii) bulk centers are not involved in the hyperfine interaction due to the spin diffusion caused by the indirect exchange interaction between the bulk centers via delocalized electrons of surface centers. These interactions fully account for the change of the characteristic spectra of polycrystalline diamond samples, observed upon a change in the average particle size from coarse (hundreds of micrometers) to fine (a few nanometers).

## Acknowledgements

The authors are grateful to M.M. Glazov for helpful discussions.